\documentclass[11pt]{revtex4}
\usepackage{amssymb,epsf}
\usepackage{latexsym}

\begin{document}

\title{Rotating Black String with Nonlinear Source}
\author{S. H. Hendi\footnote{email address: hendi@mail.yu.ac.ir}}
\affiliation{Physics Department, College of Sciences, Yasouj University, Yasouj 75914,
Iran\\
Research Institute for Astrophysics and Astronomy of Maragha (RIAAM), P.O.
Box 55134-441, Maragha, Iran\\
National Elite Foundation, P.O. Box 19615-334, Tehran, Iran}

\begin{abstract}
In this paper, we derive rotating black string solutions in the
presence of two kinds of nonlinear electromagnetic fields, so
called Born-Infeld and power Maxwell invariant. Investigation of
the solutions show that for the Born-Infeld black string the
singularity is timelike and the asymptotic behavior of the
solutions are anti-deSitter, but for power Maxwell invariant
solutions, depend on the values of nonlinearity parameter, the
singularity may be timelike as well as spacelike and the solutions
are not asymptotically anti-deSitter for all values of the
nonlinearity parameter. Next, we calculate the conserved
quantities of the solutions by using the counterterm method, and
find that these quantities do not depend on the nonlinearity
parameter. We also compute the entropy, temperature, the angular
velocity, the electric charge and the electric potential of the
solutions, in which the conserved and thermodynamics quantities
satisfy the first law of thermodynamics.
\end{abstract}

\maketitle


\section{Introduction}
The nonlinear source of general relativity attract the significant
attention because of the specific properties such as the black
hole solutions with interesting asymptotic behaviors, existence of
the soliton solutions, solving the initial singularity in the
early universe and so on. The pioneering theory of the nonlinear
electrodynamics was proposed, by Born and Infeld \cite{BI}, with
the aim of obtaining a finite value for the self-energy of a
pointlike charge. In 1935, Hoffmann \cite{Hoff} attempted to
relate the nonlinear electrodynamics and gravity. He obtained a
solution to the Einstein equations for a pointlike Born-Infeld
charge, which is devoid of the divergence of the metric at the
origin that characterizes the Reissner-Nordstr\"{o}m solution.

After these great achievements, because of difficulties that
appeared in mathematical formulation, there were not any
appreciable attentions to the nonlinear electromagnetic fields. In
other word, the Born-Infeld theory was nearly forgotten for
several decades, until the interest in nonlinear electrodynamics
increased in the context of low energy string theory \cite{Frad}.
The exact solutions of Born-Infeld theory coupled to gravity with
or without a cosmological constant have been considered by many
authors \cite{Demi}. The rotating solutions of Born-Infeld gravity
with various horizons and investigation of their properties have
been considered in Ref. \cite{BIpaper}.

In recent years there has been aroused interest about black hole
solutions whose source is Maxwell invariant raised to the power
$s$, i.e., $\left( F_{\mu \nu }F^{\mu \nu }\right) ^{s}$ as the
source of geometry in Einstein and higher derivative gravity
\cite{PMIpaper}. This theory is considerably richer than that of
the linear electromagnetic field and in the special case ($s=1$)
it can reduce to linear field. Also, it is valuable to find and
analyze the effects of exponent $s$ on the behavior of the new
solutions and the laws of black hole mechanics \cite{Rasheed}. In
addition, in higher dimensional gravity, for the special choice
$s=d/4$, where $d=$dimension of the spacetime is a multiple of
$4$, it yields a traceless Maxwell's energy-momentum tensor which
leads to conformal invariance. The idea is to take advantage of
the conformal symmetry to construct the analogues of the four
dimensional Reissner-Nordstr\"{o}m solutions, in higher dimensions
\cite{Conformalpaper}.

Being motivated by the gravity originated from the nonlinear
electromagnetic fields, in this paper we investigate the existence
of black string solutions with two types of nonlinear
electromagnetic fields, called Born-Infeld theory (BI) and a power
of Maxwell invariant (PMI), separately and study their properties.
In general, the black string solutions of Einstein gravity have
been extensively analyzed in the many literatures. For e.g.,
charge black string solutions have been studied in Ref.
\cite{Horne}, the nonuniform black strings have been considered in
\cite{Kudoh}, thermodynamics and stability of black strings have
been investigated in \cite{Gregory,Dehghani1} and also the
properties of the non-abelian and nontrivial topology black string
have been studied in \cite{Hartmann} and \cite{Kol}, respectively.

Organization of the paper is as follows: we give a brief review of
the field equations of Einstein gravity sourced by the nonlinear
electromagnetic fields, present charged rotating black string
solutions and investigate their properties, especially asymptotic
behavior of them. Then, we obtain conserved and thermodynamic
quantities of the black string in which satisfy the first law of
thermodynamics. We finish our paper with some conclusions.

\section{Rotating Black String with nonlinear electromagnetic field}
The $4$-dimensional action of Einstein gravity with nonlinear
electromagnetic field in the presence of cosmological constant is
given by
\begin{equation}
I_{G}=-\frac{1}{16\pi }\int_{\mathcal{M}}d^{4}x\sqrt{-g}\left[
R-2\Lambda +L(\mathcal{F})\right] -\frac{1}{8\pi }\int_{\partial
\mathcal{M}}d^{3}x\sqrt{-\gamma }\Theta (\gamma ),  \label{Act}
\end{equation}%
where ${R}$ is the Ricci scalar, $\Lambda $ refers to the negative
cosmological constant which in general is equal to $-3/l^{2}$ for
asymptotically anti-deSitter solutions, in which $l$ is a scale length
factor.

In Eq. (\ref{Act}), $L(\mathcal{F})$ is the Lagrangian of nonlinear
electromagnetic field. Here we consider two classes of nonlinear
electromagnetic fields, namely Born-Infeld (BI) and power of Maxwell
invariant (PMI) in which their Lagrangians are
\begin{equation}
L(\mathcal{F})=\left\{
\begin{array}{cc}
4\beta ^{2}\left( 1-\sqrt{1+\frac{\mathcal{F}}{2\beta
^{2}}}\right) & ,\text{BI} \\
-\left( \alpha \mathcal{F}\right) ^{s} & ,\text{PMI}%
\end{array}%
\right. .  \label{LagEM}
\end{equation}%
In this equation, $\beta $ is called the Born-Infeld parameter
with dimension of mass, the exponent $s$ is related to the power
of nonlinearity, the Maxwell invariant $\mathcal{F}=F_{\mu \nu
}F^{\mu \nu }$ in which $F_{\mu \nu }=\partial _{\mu }A_{\nu
}-\partial _{\nu }A_{\mu }$ is the electromagnetic field tensor
and $A_{\mu }$ is the gauge potential. $L(\mathcal{F})$ reduces to
the standard Maxwell form $L(\mathcal{F})=-\mathcal{F}$, when
$\beta \rightarrow \infty $ and $s\rightarrow 1$ for BI and PMI
theory, respectively.

The last term in Eq. (\ref{Act}) is the Gibbons-Hawking surface
term. It is required for the variational principle to be
well-defined. The factor $\gamma $\ and $\Theta$ are,
respectively, the trace of the induced metric and extrinsic
curvature for the boundary ${\partial \mathcal{M}}$. Varying the
action (\ref{Act}) with respect to the gravitational field $g_{\mu
\nu }$ and the gauge field $A_{\mu }$, the field equations are
obtained as
\begin{equation}
R_{\mu \nu }-\frac{1}{2}g_{\mu \nu }\left( R-2\Lambda \right) =\alpha T_{\mu
\nu },  \label{FE1}
\end{equation}
\begin{equation}
\partial _{\mu }\left(\sqrt{-g} L^{\prime }(\mathcal{F}) F^{\mu \nu }\right)
=0,  \label{FE2}
\end{equation}
where
\begin{equation}
T_{\mu \nu }=\frac{1}{2}g_{\mu \nu }L(\mathcal{F})-2F_{\mu \lambda
}F_{\nu}^{\; \lambda }L^{\prime }(\mathcal{F}),  \label{FE3}
\end{equation}
and $L^{\prime }(\mathcal{F})=\frac{dL(\mathcal{F})}{d\mathcal{F}}$. Our
main aim here is to obtain charged rotating black string solutions of the
field equations (\ref{FE1}) - (\ref{FE3}) and investigate their properties.
We assume the rotating metric has the following form \cite{Lem}
\begin{equation}
ds^{2}=-f(r)\left( \Xi dt-ad\phi \right) ^{2}+\frac{r^{2}}{l^4}\left( a
dt-\Xi l^2 d\phi \right) ^{2}+\frac{dr^{2}}{f(r)}+\frac{r^{2}}{l^{2}}dz^{2},
\label{Metric}
\end{equation}
where $\Xi =\sqrt{1+a^{2}/l^{2}}$, $a$ is the rotation parameter
and the functions $f(r)$ should be determined. The two dimensional
space, $t$ =constant and $r$ =constant, has the topology $R\times
S^{1}$, and $0\leq \phi <2\pi $, $-\infty <z<\infty $. It is easy
to show that for the metric (\ref{Metric}), the Kretschmann scalar
is
\begin{equation}
R_{\mu \nu \rho \sigma }R^{\mu \nu \rho \sigma }=f^{\prime \prime
2}(r)+\left( \frac{2f^{\prime }(r)}{r}\right) ^{2}+\left( \frac{2f(r)}{r^{2}}
\right) ^{2},  \label{RR}
\end{equation}
where prime and double primes denote first and second derivative
with respect to $r$, respectively. Also one can show that other
curvature invariants (such as Ricci scalar, Ricci square, Weyl
square and so on) are functions of $f^{\prime \prime }$,
$f^{\prime }/r$ and $f/r^{2}$ and therefore it is sufficient to
study the Kretschmann scalar for the investigation of the
spacetime curvature.

Here, we use the gauge potential ansatz
\begin{equation}
A_{\mu }=h(r)\left( \Xi \delta _{\mu }^{0}-a \delta _{\mu }^{\phi
}\right) \label{Amu}
\end{equation}%
in the nonlinear electromagnetic fields equation (\ref{FE2}). We obtain
\begin{equation}
h(r)=\left\{
\begin{array}{cc}
-\frac{q}{r}\ \times {_{2}F_{1}\left( \left[
\frac{1}{2},\frac{1}{4}\right] , \left[ \frac{5}{4}\right]
,-\frac{{q^{2}}}{\beta ^{2}r^{4}}\right) } & ,
\text{BI} \\
\left\{
\begin{array}{cc}
q\ln r, & s=\frac{3}{2} \\
-qr^{\lambda }, & s\neq \frac{3}{2}
\end{array}
\right. & ,\text{PMI}
\end{array}
\right.  \label{h(r)}
\end{equation}
where $q$ is an integration constant which is related to the
electric charge of black string, $_{2}F_{1}([a,b],[c],z)$ is
hypergeometric function and $\lambda =(2s-3)/(2s-1)$. It is
notable that for $s=0,1/2$, the function $h(r)=Constant$, we do
not have electromagnetic field and therefore in the rest of the
presented paper we exclude these cases from our discussion.
\begin{figure}[tbp]
\epsfxsize=10cm \centerline{\epsffile{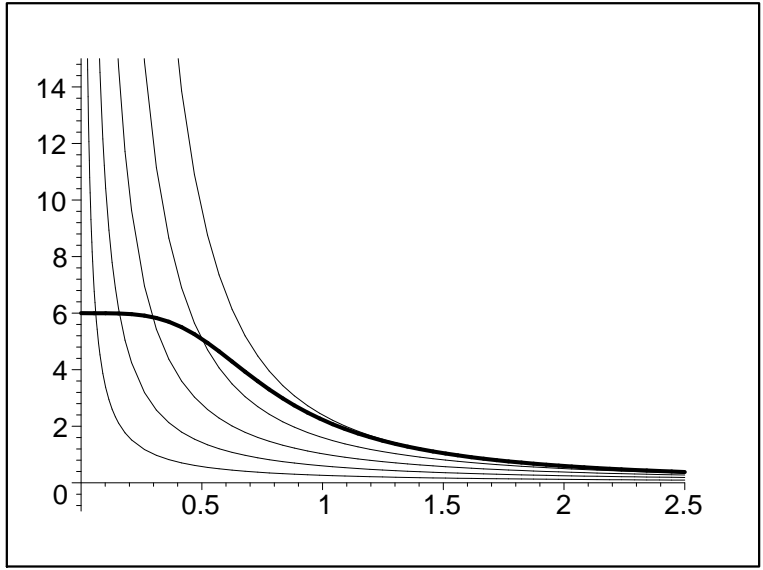}}
\caption{$F_{tr}$ versus $r$ for $\Xi =1.2$, $q=2$, $\protect\beta =5$ ( for
BI theory: Bold line) and $s=1$, $s=1.1$, $s=1.2$, $s=1.3$ and $s=1.4$ ( for
PMI theory: from right to left, respectively).}
\label{Ftr}
\end{figure}
One may note that in the linear limit, $\beta \rightarrow \infty $ for BI
branch and also $s=1$ for PMI branch, $A_{\mu }$ of Eq. (\ref{Amu}) reduces
to the gauge potential of linear Maxwell field \cite{Dehghani2}. It is easy
to show that the non-vanishing components of the electromagnetic field
tensor can be written in the form
\begin{eqnarray}
F_{tr} &=&-\frac{\Xi q}{r}\times \left\{
\begin{array}{cc}
-\beta r\left( \beta ^{2}r^{4}+{q^{2}}\right) ^{-1/2} & ,\text{BI} \\
\left\{
\begin{array}{cc}
-1, & s=\frac{3}{2} \\
\lambda r^{\lambda }, & s\neq \frac{3}{2}
\end{array}
\right. & ,\text{PMI}
\end{array}
\right. , \\
F_{\phi r} &=&-\frac{a}{\Xi }F_{tr}.
\end{eqnarray}
It is remarkable that for $s=3/2$, the electromagnetic field is
proportional to $r^{-1}$, in which the same as charged BTZ black
hole \cite{BTZ,BTZlike}. Also, it is interesting to note that, in
general case, the expression of the electric field depends on the
nonlinearity parameter ($\beta $ or $s$), and its value coincides
with the $4$-dimensional Reissner-Nordstr\"{o}m solutions for the
linear limit. We can see from Fig. (\ref{Ftr}), the
electromagnetic field vanishes at large values of $r$, as it
should be. But near the origin, in contrast with the finite value
of the BI theory, the electromagnetic field diverges for PMI
theory. In addition, this figure shows that the nonlinearity
parameter, $s$, affects on the strength of divergency for
$r\rightarrow 0$.

Now, we should fix the sign of the constant $\alpha $ in order to ensure the
real solutions. It is easy to show that
\[
\mathcal{F}=F_{\mu \nu }F^{\mu \nu }=-2\left[ h^{\prime }(r)\right] ^{2},
\]
and so the power Maxwell invariant, $\left( \alpha \mathcal{F}\right) ^{s}$,
may be imaginary for positive $\alpha $, when $s$ is fractional. Therefore
we set $\alpha =-1$, to have real solutions without loss of generality.

To find the metric function $f(r)$, one may use any components of
Eq. (\ref{FE1}). The simplest equation is the $rr$ component of
these equations, which can be written as
\begin{equation}
2rf^{^{\prime }}(r)+2\Lambda r^{2}+2f(r)+\Upsilon (r)=0,  \label{rrcomp}
\end{equation}
where
\[
\Upsilon (r)=\left\{
\begin{array}{cc}
4{\beta }\left[ \left( \beta ^{2}r^{4}+{q^{2}}\right) ^{1/2}-\beta
r^{2}
\right]  & ,\text{BI} \\
-r^{2}(2s-1)\left[ 2h^{^{\prime }2}(r)\right] ^{s} & ,\text{PMI}
\end{array}
\right. .
\]
The solutions of Eq. (\ref{rrcomp}) can be written as
\begin{equation}
f(r)=-\frac{\Lambda r^{2}}{3}-{\frac{m}{r}+}\left\{
\begin{array}{cc}
\begin{array}{c}
{\frac{2{\beta }}{3}}\left[ \beta r^{2}-\left( \beta
^{2}r^{4}+{q^{2}}
\right) ^{1/2}\right] + \\
{\frac{4q^{2}}{3r^{2}}}\ _{2}F_{1}{\left( \left[
\frac{1}{2},\frac{1}{4} \right] ,\left[ \frac{5}{4}\right]
,-\frac{{q^{2}}}{\beta ^{2}r^{4}}\right) }
\end{array}
& ,\text{BI} \\
&  \\
\left\{
\begin{array}{cc}
2^{s-1}\left( 2s-1\right) \lambda ^{2s-1}q^{2s}r^{\lambda -1}, &
0<s<\frac{1}{2},s<0 \\
\frac{-2^{3/2}q^{3}\ln r}{r}, & s=\frac{3}{2} \\
-2^{s-1}\left( 2s-1\right) \lambda ^{2s-1}q^{2s}r^{\lambda -1}, &
\text{ Otherwise}
\end{array}
\right.  & ,\text{PMI}
\end{array}%
\right. ,  \label{F(r)}
\end{equation}
where $m$ is the integration constant which is related to mass
parameter. One can check that the solutions given by Eq.
(\ref{F(r)}) satisfy all the components of the field equations
(\ref{FE1}). It is notable that for $s=3/2$, the charge term in
(\ref{F(r)}) includes logarithmic term and is different from other
cases. This special solution ($s=3/2$), is very close to BTZ
solutions (see \cite{BTZlike} for more details).
\begin{figure}[tbp]
\epsfxsize=6cm \centerline{\epsffile{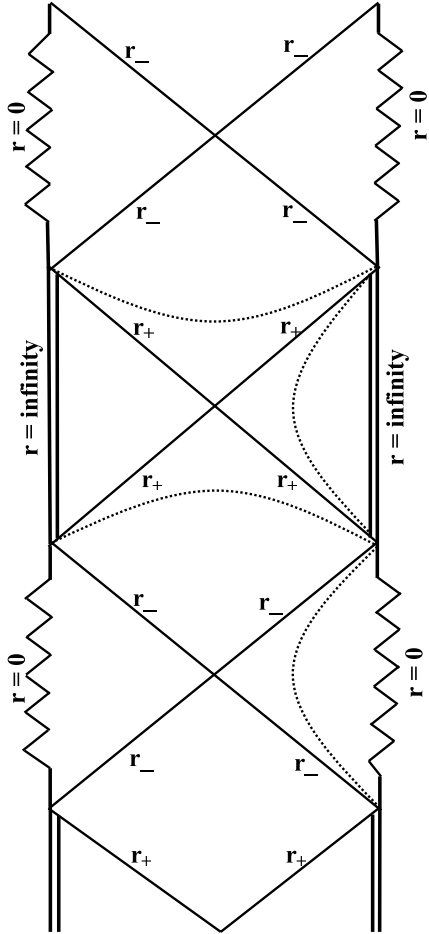}}
\caption{Penrose diagram with negative $\Lambda $ for BI theory with
arbitrary $\protect\beta $ and PMI theory when $0<s<\frac{3}{2}$, with two
horizons $r_{+}$ and $r_{-}$ and timelike singularity at $r=0$ in which
dotted curves represent $r=constant$.}
\label{pen1}
\end{figure}
\begin{figure}[tbp]
\epsfxsize=4cm \centerline{\epsffile{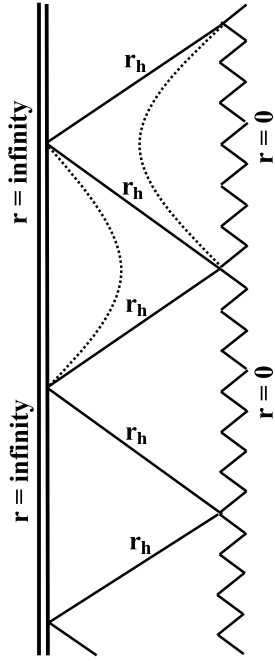}}
\caption{Penrose diagram with negative $\Lambda $ for BI theory with
arbitrary $\protect\beta $ and PMI theory when $0<s<\frac{3}{2}$, with one
horizon (extreme black string) $r_{+}=r_{-}=r_{h}$ and timelike singularity
at $r=0$ in which dotted curves represent $r=constant$.}
\label{pen2}
\end{figure}
\begin{figure}[tbp]
\epsfxsize=6cm \centerline{\epsffile{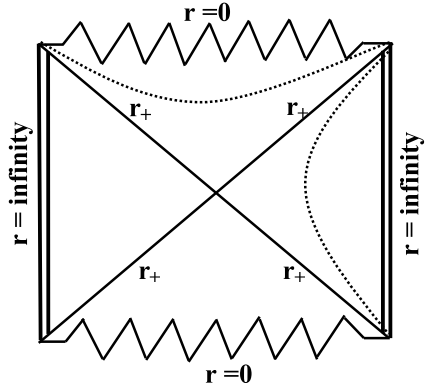}}
\caption{Penrose diagram with negative $\Lambda $ for PMI theory when $s<0$
or $s\geq \frac{3}{2}$, with one horizon $r_{+}$ and spacelike singularity
at $r=0$ in which dotted curves represent $r=constant$.}
\label{pen3}
\end{figure}
\begin{figure}[tbp]
\epsfxsize=10cm \centerline{\epsffile{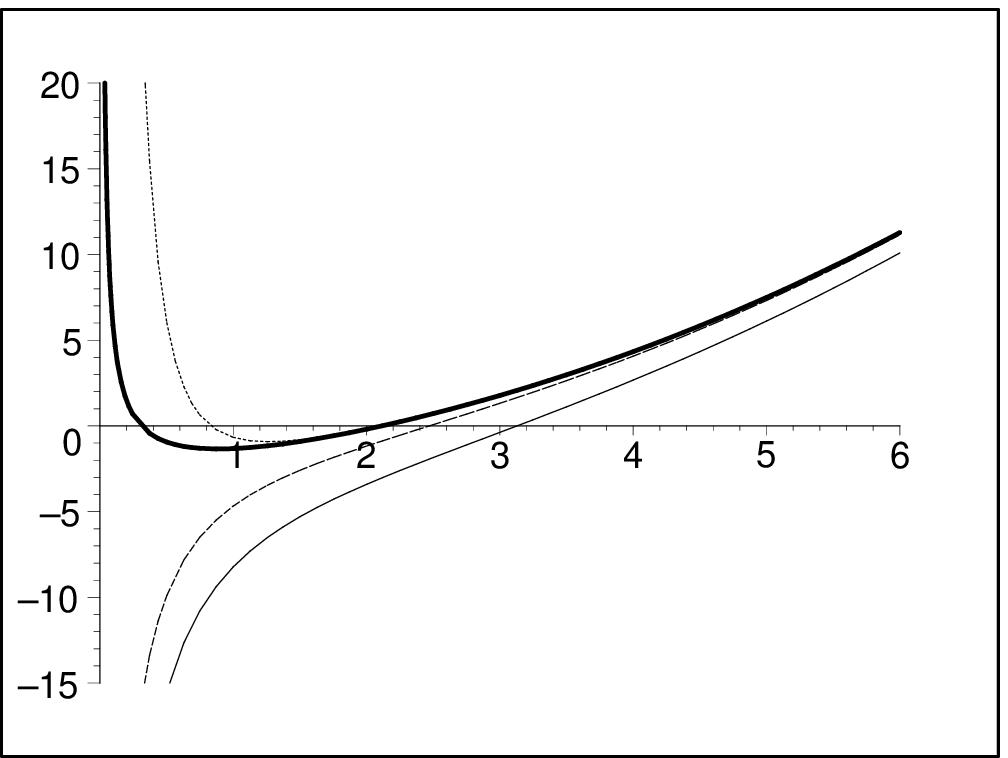}} \caption{$f(r)$
versus $r$ for $\Lambda =-1$, $q=2$, $m=5$, $\protect\beta =0.7$
for BI theory (Bold line) and $s=1$ (dotted line), $s=2$ (solid
line), $s=-2$ (dashed line) for PMI theory.} \label{f(r)}
\end{figure}

\subsection{Properties of the solutions}
The metric function $f(r)$, presented here, differs from the
linear $4$ -dimensional Reissner-Nordstr\"{o}m black hole
solutions; it is notable that the electric charge term in the
linear case is proportional to $r^{-2}$, but in the presented
metric function, this term depends on the nonlinearity parameter
($\beta $ or $s$). It is notable that in the linear case ($\beta
\rightarrow \infty $ for BI or $s=1$ for PMI), the presented
solutions reduce to the asymptotically anti-deSitter charged
rotating black string \cite{Lem}.

In order to study the general structure of these spacetime, we
first explain the crucial role of negative cosmological constant.
One can find that for vanishing cosmological constant, we have
physical solutions only for $s=1$ in PMI source and $\beta
\rightarrow \infty$ for BI theory (linear source). In other word,
for nonlinear sources we need to insert negative cosmological
constant in the gravitational field equations to obtain meaningful
solutions. This is due to the fact that, repulsive gravitational
contribution, arising from nonlinear electromagnetic source,
should balance by the attractive contribution of the negative
cosmological constant (for more details see \cite{CC}).

existence of nonlinear source, the black string solutions exist
only in the presence of negative cosmological constant, when the
repulsive gravitational contribution, originate from nonlinear
electromagnetic source, is balanced by the attraction of the
negative cosmological constant.

Second, we investigate the effects of the nonlinear
electromagnetic field on the asymptotic behavior of the solutions.
The solution of BI theory is asymptotically anti-deSitter for all
values of nonlinearity parameter $\beta $. But in PMI solutions
the asymptotic behaviors are different. It is worthwhile to
mention that for $0<s<\frac{1}{2 }$, the asymptotic dominant term
of Eq. (\ref{F(r)}) is third term and the presented solutions are
not asymptotically anti-deSitter, but for the cases $s<0$\ or
$s>\frac{1}{2}$ (include of $s=\frac{3}{2}$), the asymptotic
behavior of rotating black string solutions are the same as linear
anti-deSitter case \cite{Lem}. Third, we look for the essential
singularity(ies). After some algebraic manipulation, one can show
that the Kretschmann scalar (\ref{RR}) with metric function
(\ref{F(r)}) diverges at $r=0$ and is finite for $r\neq 0$. Thus,
there is a curvature singularity located at $r=0$. In addition,
Penrose diagrams, Figs. (\ref{pen1})-(\ref{pen3}) and also Fig.
(\ref{f(r)}) show that the singularity is timelike for BI theory
and $0<s<\frac{3}{2}$ in PMI theory, but for $s<0$ or $s \geq
\frac{3}{2}$ it is spacelike. Drawing the Penrose diagrams show
that the casual structure of the solutions are asymptotically well
behaved.

\subsection{Conserved quantities \label{Therm}}
Next, we calculate the conserved quantities of the solutions. To
compute the conserved charges of our solutions, we use the
approach proposed by Balasubramanian and Kraus in \cite{Kraus}.
This technique was inspired by \emph{anti-deSitter/conformal field
theory} correspondence \cite{Mal} and consists in adding suitable
counterterms $I_{ct}$ to the action of the theory (\ref{Act}) in
order to ensure the finiteness of the boundary stress tensor
$T_{ab}=\frac{2}{\sqrt{-\gamma }}\frac{\delta I}{\delta \gamma
^{ab}}$ derived by the quasilocal energy definition \cite{BY}.
Therefore we supplement the general action (\ref{Act}) with the
following boundary counterterm
\begin{equation}
I_{ct}=-\frac{1}{4\pi }\int_{\partial
\mathcal{M}}d^{3}x\sqrt{-\gamma } \left( -\frac{1}{l}\right) .
\label{cont}
\end{equation}
Varying the total action ($I_{tot}=I_{G}+I_{ct}$) with respect to the
induced metric $\gamma _{ab}$, we find the divergence-free boundary
stress-tensor
\begin{equation}
T^{ab}=\frac{\Theta ^{ab}-\left( \Theta +\frac{2}{l}\right) \gamma
^{ab}}{8\pi },  \label{Stres}
\end{equation}
To compute the conserved charges of the spacetime, one should
choose a spacelike surface $\mathcal{B}$ in $\partial \mathcal{M}$
with metric $\sigma _{ij}$, and write the boundary metric in
Arnowitt-Deser-Misner form
\[
\gamma _{ab}dx^{a}dx^{a}=-N^{2}dt^{2}+\sigma _{ij}\left( d\varphi
^{i}+V^{i}dt\right) \left( d\varphi ^{j}+V^{j}dt\right) ,
\]
where the coordinates $\varphi ^{i}$ are the angular variables
parameterizing the hypersurface of constant $r$ around the origin,
and $N$ and $V^{i}$ are the lapse and shift functions,
respectively. When there is a Killing vector field $\mathcal{\xi
}$ on the boundary, the quasilocal conserved quantity associated
with the stress tensor of Eq. (\ref{Stres}) can be written as
\begin{equation}
Q(\mathcal{\xi )}=\int_{\mathcal{B}}d^{2}x\sqrt{\sigma
}T_{ab}n^{a}\mathcal{\ \xi }^{b},  \label{charge}
\end{equation}
where $\sigma $ is the determinant of the metric $\sigma _{ij}$,
$\mathcal{\ \xi }$ and $n^{a}$ are, respectively, the Killing
vector field and the unit normal vector on the boundary
$\mathcal{B}$. In the context of counterterm method, the limit in
which the boundary $\mathcal{B}$ becomes infinite
($\mathcal{B}_{\infty }$) is taken, and the counterterm
prescription ensures that the total action and conserved charges
are finite \cite{Kraus}. For boundaries with timelike ($\xi
=\partial /\partial t$) and rotational ($\varsigma =\partial
/\partial \varphi $) Killing vector fields, one obtains the
quasilocal mass and angular momentum
\begin{eqnarray}
M &=&\int_{\mathcal{B}}d^{2}x\sqrt{\sigma }T_{ab}n^{a}\xi ^{b},
\label{Mastot} \\
J &=&\int_{\mathcal{B}}d^{2}x\sqrt{\sigma }T_{ab}n^{a}\varsigma ^{b}.
\label{Angtot}
\end{eqnarray}
These quantities are, respectively, the conserved mass and angular
momenta of the system enclosed by the boundary $\mathcal{B}$. The
mass and angular momentum per unit length of the string when the
boundary $\mathcal{B}$ goes to infinity can be calculated through
the use of Eqs. (\ref{Mastot}) and (\ref{Angtot}). We find
\[
{M}=\frac{1}{16\pi l}\left( 3\Xi ^{2}-1\right) m,
\]
\[
{J}=\frac{3}{16\pi l}\Xi ma.
\]
For $a=0$ ($\Xi =1$), the angular momentum per unit length
vanishes, and therefore $a$ is the rotational parameter of the
spacetime.

Next, we obtain the entropy of the black string. Bekenstein argued
that the entropy of a black hole is a linear function of the area
of its event horizon, which so-called area law \cite{Bekenstein}
and also he proposed a value for the proportionality constant,
deduced from a semiclassical calculation of the minimum increase
in the area of a black hole when it absorbs a particle. Since the
area law of the entropy is universal, and applies to all kinds of
black objects in Einstein gravity \cite{Bekenstein,hunt},
therefore the entropy per unit length of the black string is
\begin{equation}
{S}=\frac{r_{+}^{2}\Xi }{4l}.  \label{ent}
\end{equation}
Although our solution is not static, the Killing vector
\begin{equation}
\chi =\partial _{t}+\Omega \partial _{\phi },  \label{Kil}
\end{equation}
is the null generator of the event horizon where $\Omega $ is the
angular velocity of the outer horizon. By analytic continuation of
the metric we can obtain the temperature and angular velocity of
the horizon. The analytical continuation of the Lorentzian metric
by $t\rightarrow i\tau $ and $a\rightarrow ia$ yields the
Euclidean section, whose regularity at $r=r_{+}$ requires that we
should identify $\tau \sim \tau +\beta _{+}$ and $\phi \sim \phi
+i\Omega \beta _{+}$, where $\beta _{+}$ is the inverse Hawking
temperature of the event horizon. We find
\begin{equation}
T_{+}=\beta _{+}^{-1}=\frac{-\Lambda r_{+}}{4\pi \Xi
}+\frac{1}{4\pi \Xi } \times \left\{
\begin{array}{cc}
2\beta ^{2}r_{+}{\left[ 1-\left( 1+\frac{{q^{2}}}{\beta
^{2}r_{+}^{4}}
\right) ^{1/2}\right] } & ,\text{BI} \\
&  \\
\left\{
\begin{array}{cc}
\frac{2^{s-1}(2s-1)\lambda ^{2s}q^{2s}}{r_{+}^{(2s+1)/(2s-1)}} &
0<s<\frac{1
}{2}\;,s<0 \\
\frac{2^{3/2}q^{3}}{r_{+}^{2}}, & s=\frac{3}{2} \\
-\frac{2^{s-1}(2s-1)\lambda ^{2s}q^{2s}}{r_{+}^{(2s+1)/(2s-1)}}, &
\text{ Otherwise}
\end{array}
\right.  & ,\text{PMI}
\end{array}
\right. ,  \label{T}
\end{equation}
and
\begin{equation}
\Omega =\frac{a}{\Xi l^{2}}.  \label{Omega}
\end{equation}
The next quantity we are going to calculate is the electric charge of the
black string. The electric charge per unit length of it, $Q$, can be found
by calculating the flux of the electromagnetic field at infinity, yielding
\begin{equation}
{Q}=\frac{\Xi }{8\pi }\times \left\{
\begin{array}{cc}
2q & ,\text{BI} \\
\left\{
\begin{array}{cc}
2^{3/2}q^{2}, & s=\frac{3}{2} \\
-2^{s}\lambda ^{2s-1}q^{2s-1}, & s\neq \frac{3}{2}
\end{array}
\right.  & ,\text{PMI}
\end{array}
\right.   \label{Q}
\end{equation}
The electric potential $U$, measured at infinity with respect to the event
horizon $r_{+}$, is defined by \cite{Cvetic}
\begin{equation}
U=A_{\mu }\chi ^{\mu }\left\vert _{r\rightarrow \infty }-A_{\mu }\chi ^{\mu
}\right\vert _{r=r_{+}},  \label{U}
\end{equation}
where $\chi $ is the null generator of the event horizon (\ref{Kil}). One
can easily obtain the electric potential as
\begin{equation}
U=\frac{q}{\Xi r_{+}}\times \left\{
\begin{array}{cc}
\ {_{2}F_{1}\left( \left[ \frac{1}{2},\frac{1}{4}\right] ,\left[
\frac{5}{4}
\right] ,-\frac{{q^{2}}}{\beta ^{2}r_{+}^{4}}\right) } & ,\text{BI} \\
\left\{
\begin{array}{cc}
-r_{+}\ln r_{+}, & s=\frac{3}{2} \\
r_{+}^{\lambda +1}, & s\neq \frac{3}{2}
\end{array}
\right.  & ,\text{PMI}
\end{array}
\right. .  \label{UU}
\end{equation}
Having the conserved and thermodynamic quantities of the rotating
black string at hand, we are in a position to check the first law
of black hole thermodynamics. Using the Smarr-type formula for
both branches, it is straightforward to calculate the temperature,
angular velocity and electric potential in the following manner
\begin{equation}
T=\left( \frac{\partial M}{\partial S}\right) _{J,Q},\text{ \ \ \
\ \ \ } \Omega =\left( \frac{\partial M}{\partial J}\right)
_{S,Q},\text{ \ \ \ \ \ \ }U=\left( \frac{\partial M}{\partial
Q}\right) _{S,J}  \label{TOmegaU}
\end{equation}
Since the quantities calculated by Eq. (\ref{TOmegaU}) coincide
with Eqs. (\ref{T}), (\ref{Omega}) and (\ref{UU}), one can
conclude that these quantities satisfy the first law of
thermodynamics
\begin{equation}
dM=TdS+\Omega d{J}+Ud{Q}.  \label{firstLaw}
\end{equation}

\section{Conclusion}
In this paper, we obtained a new class of rotating black string
solutions in the presence of negative cosmological constant and
investigated their properties. The matter fields, in which we
considered, are two kinds of nonlinear electromagnetic fields,
called Born-Infeld theory and power Maxwell invariant source. As
expected, the nonlinearity parameters effected on the
electromagnetic field, clearly. Although investigation of
nonlinear electromagnetic source is more complicated, but
nonlinear sources are more flexible. For e.g., it is interesting
that we could obtain the BTZ-like solution for $s=3/2$.

In addition, it is worthwhile to mention that the repulsive
gravitational contribution arising from nonlinear sources is
balanced by the attractive contribution of the negative
cosmological constant. In other words, we did not encounter with
physical solutions if we considered the nonlinear sources without
$\Lambda$ term.

After presented physical solutions, we investigated their
properties. As one can see from Penrose diagrams, the singularity
of the Born-Infeld black string is timelike and the asymptotic
behavior of the BI solution is anti-deSitter, but for power
Maxwell invariant source, the singularity type and the asymptotic
behavior of the solutions depend on the values of nonlinearity
parameter.

Then by using the counterterm approach, we calculated the
conserved quantities of the solutions which they did not change
with respect to the linear Maxwell field. Also, using the area
law, Gauss law and analytic continuation of the metric, we
achieved the entropy, the electric charge and the temperature and
angular velocity of the solutions. It is notable that, in contrast
with the Born-Infeld black string, the nonlinearity parameter of
the power Maxwell invariant solutions effected on the electrical
charge. Finally, we obtained the electric potential of the black
string and checked that these conserved and thermodynamic
quantities satisfy the first law of thermodynamics.

\acknowledgments{ This work has been supported financially by
Research Institute for Astronomy and Astrophysics of Maragha.}


\end{document}